\documentclass[aps,pra,twocolumn,groupedaddress,showpacs,superscriptaddress,amssymb,amsmath]{revtex4-1}
\usepackage{graphicx}
\usepackage{tabularx}
\usepackage{color}
\usepackage{amsmath}
\usepackage{comment}
\newcommand{\be}{\begin{equation}}
\newcommand{\ee}{\end{equation}}
\newcommand{\bea}{\begin{eqnarray}}
\newcommand{\eea}{\end{eqnarray}}
\usepackage{dcolumn}
\usepackage{hyperref}
\usepackage{bm}
\usepackage{epsf}
\usepackage{subfigure}
\usepackage{epstopdf}%
\usepackage{amsfonts}

\begin{document}

\title
{The Anderson impurity model out-of-equilibrium: 
Assessing the accuracy of simulation techniques with an exact current-occupation relation}

\author{Bijay Kumar Agarwalla}
\affiliation{Chemical Physics Theory Group, Department of Chemistry, and
Centre for Quantum Information and Quantum Control,
University of Toronto, 80 Saint George St., Toronto, Ontario, Canada M5S 3H6}
\author{Dvira Segal}
\affiliation{Chemical Physics Theory Group, Department of Chemistry, and
Centre for Quantum Information and Quantum Control,
University of Toronto, 80 Saint George St., Toronto, Ontario, Canada M5S 3H6}

\date{\today}

\begin{abstract}
We study the interacting, symmetrically coupled single impurity Anderson model.
By employing the nonequilibrium Green's function formalism,
we establish an exact relationship between the 
steady-state charge current flowing through the impurity (dot) and its occupation.
We argue that the steady state current-occupation relation can be used to assess the consistency 
of simulation techniques, and identify spurious transport phenomena.
We test this relation in two different model variants:
First, we study the Anderson-Holstein model in the strong electron-vibration coupling limit 
using the polaronic quantum master equation method. 
We find that the current-occupation relation
is violated numerically in standard calculations, with simulations bringing up incorrect
transport effects. Using a numerical procedure, we resolve the problem efficiently.
Second, we simulate the Anderson model with electron-electron interaction on the dot
using a deterministic numerically-exact time-evolution scheme. 
Here, we observe that the current-occupation relation is satisfied in the steady-state limit---even before results
converge to the exact limit.
%
%
\end{abstract}
\maketitle

\section{Introduction}
\label{Sintro}

The single impurity Anderson model \cite{Anderson} plays a central role 
in many-body condensed-phases physics. While it was originally introduced for describing the
thermodynamics of magnetic impurities in non-magnetic metals \cite{Kondo}, it has
received significant attention more recently in the context of transport at the nanoscale \cite{Galperin-review}.
In the standard nonequilibrium setup, a single electronic level (impurity, dot) with two interacting electrons
is coupled to two voltage-biased electronic reservoirs (metals).
This simple scenario can be made more involved by including the interaction of electrons with 
local vibrations (``Anderson-Holstein model"), and by considering multiple electronic states \cite{Galperin-review}.

The Anderson impurity model had greatly contributed to the 
understanding of charge and energy transport mechanisms        
in nanoscale systems, quantum dots and molecular junctions, by
revealing rich dynamical and steady state behavior, see e.g.  Ref. \cite{GalperinS}.
Moreover, the model serves as a playground for developing and benchmarking 
approximation schemes and simulation techniques; 
a very partial list includes 
renormalization group techniques \cite{Wilson, RG}, Green's function approaches \cite{MW93,Galperin-review},
the multilayer multiconfiguration time-dependent Hartree method \cite{Wang1,Wang2,Wang3,Rabani14}, 
Hierarchical equations of motion \cite{HEOM1,HEOM2,HEOM3},
Quantum Monte Carlo \cite{Lothar,Coheninch,CohenHT} and influence-functional path integral tools \cite{egger1,egger2,IF1,IF2}.

Beyond the study of nonlinear transport characteristics, the search for universal relations between 
different nonequilibrium observables has been central to nonequilibrium physics. 
For example, linear irreversible thermodynamics is summarized in terms of the universal Green-Kubo relations, 
with the conductance expressed in terms of an equilibrium density correlation function. 
Onsager's reciprocity theorem relates cross-transport coefficients. 
Moreover, in the past two decades various universal relations have been discovered for out-of-equilibrium systems,
now collectively referred to as ``fluctuation symmetries" \cite{fluc1,fluc2}. 


In this paper, we discuss an exact relation between the steady state charge current and the average 
charge occupation for the interacting, symmetrically coupled single impurity model. 
We argue that this relationship can be used to assess the consistency and accuracy of approximation techniques,
and in particular identify spurious transport effects.
We test this current-occupation relationship in two models: 
(i) the single-electron Anderson-Holstein model, by using a polaronic quantum master equation (QME), 
and (ii) the electron-electron interacting Anderson model, by employing a numerically exact time evolution scheme.
As we show below, in both cases the violation/verification of the 
current-occupation relationship reveals important information 
on the performance (consistency, accuracy) of simulation techniques.

The paper is organized as follows. 
In Sec.~\ref{SModel}, we present the model Hamiltonian and derive the exact relation between 
the charge current and the average occupation on the dot. 
We test this relation in Sec. \ref{SAH}, 
for the Anderson-Holstein model, and in Sec. \ref{SINFPI}, for the case with electron interaction on the dot.
We conclude in Sec. \ref{Ssum}.


\section{Exact relation between current and occupation}
\label{SModel}

We are interested in the steady state behavior of the Anderson impurity model. The total Hamiltonian
$H=H_0+H_n$ includes the noninteracting (quadratic) term $H_0$ and the many-body 
(nonlinear) contribution $H_n$. Here,
\be
H_0= H_{d0} + H_L + H_R + H_{dL} + H_{dR},
\ee 
where  $H_{d0}=\epsilon_d d^{\dagger}d$ comprises a single electronic site (dot) with
creation and annihilation operators $d^{\dagger}$, $d$, respectively.
The dot is placed in contact with two non-interacting fermionic environments (metal leads), $\alpha=L,R$, 
$H_{\alpha}= \sum_{k} \epsilon_{k \alpha} c^{\dagger}_{k \alpha} c_{k \alpha}$, 
where $k$ is the index for the electron momentum
and $c_{k \alpha}^{\dagger}$ ($c_{k \alpha})$ is the creation (annihilation) operator in the lead $\alpha$. 
Electron hopping between the dot and the leads is described by the standard tunnelling Hamiltonian
$H_{d \alpha}= \sum_{k} v_{k \alpha} c_{k\alpha}^{\dagger} d + h.c.$, 
with $v_{k\alpha}$ as the metal-dot coupling energy.
$H_n$ includes many-body interactions---assumed to affect or
couple to the $d$, $d^{\dagger}$ operators. 
For example, $H_n$ may comprise electron-electron (e-e) repulsion interaction on the dot (generalizing then
$H_0$ to include two spin species). As well, $H_n$ may
collect additional degrees of freedom, phonons and photons,
and their interaction with electrons on the dot.

The two metal leads are maintained out of equilibrium by introducing different chemical potentials on the leads, 
$\mu_L\neq \mu_R$, and different temperatures, $T_L\neq T_R$ (in our simulations below we assume equal temperatures).
The resulting steady state charge current, say from the $L$ lead towards the dot, 
is given by the celebrated Meir-Wingreen formula  \cite{meir-wingreen,Haug-book} (we assume $\hbar=1$),
\be
\langle I_L \rangle = e \int_{-\infty}^{\infty} \frac{d\epsilon}{2 \pi} \, 
\big[G_d^{>}(\epsilon) \,\Sigma_L^{<}(\epsilon) - G_d^{<}(\epsilon) \, \Sigma_L^{>}(\epsilon) \big].
\label{eq:MW}
\ee 
This formal expression holds for an arbitrarily-interacting impurity. 
Here, the principal object is the one-electron interacting Green's function
$G_d(\tau, \tau')=-i\langle {\rm T_c}d(\tau)d^{\dagger}(\tau') \rangle$, defined
on the Keldysh contour  with $T_c$ as the contour-ordering operator
\cite{Haug-book, Rammer_review, Rammer_book}.
Projecting to real time, then to the energy  domain,
$G_{d}^{</>}(\epsilon)$ are the lesser and greater components of the one-electron Green's function.
$\Sigma_{L,R}^{</>}(\epsilon)$ are the corresponding self-energy terms, 
responsible for electron transfer in and out of the central system to the $L$ and $R$ metal leads.
These terms can be analytically obtained \cite{Haug-book, Rammer_review, Rammer_book},
\bea
\Sigma_{\alpha}^{<}(\epsilon) = i f_{\alpha}(\epsilon) \Gamma_{\alpha}(\epsilon),  \,\,\,\,
\Sigma_{\alpha}^{>}(\epsilon) = -i \big(1- f_{\alpha}(\epsilon)\big) \Gamma_{\alpha}(\epsilon),
\nonumber\\
\eea
with $\Gamma_{\alpha}(\epsilon)= 2 \pi \sum_{k} |v_{k\alpha }|^2 \delta (\epsilon-\epsilon_{k\alpha})$ as the impurity-lead
hybridization energy.
$f_{\alpha}(\epsilon)=1/[e^{\beta_{\alpha} (\epsilon-\mu_{\alpha})}+1]$ is the Fermi-Dirac 
distribution function for the $\alpha$ lead, $\beta_{\alpha}=1/T_{\alpha}$ is the inverse temperature, $k_B\equiv 1$.

From here on we assume that the impurity is symmetrically coupled to the two leads, 
$\Gamma(\epsilon)\equiv \Gamma_{\alpha}(\epsilon)$. 
Eq.~(\ref{eq:MW}) then reduces to the simple form
\be
\langle I_L \rangle  = e \int_{-\infty}^{\infty} \frac{d\epsilon}{4 \pi } \, \big[f_L(\epsilon)-f_R(\epsilon)\big] \,\Gamma(\epsilon) \,A(\epsilon; \{\mu_{\alpha},T_{\alpha}\}),
\label{eq:Lan-cur}
\ee
with
\bea
A(\epsilon; \{\mu_{\alpha},T_{\alpha}\}) = i \left[G_d^r(\epsilon)-G_d^a(\epsilon)\right],
\eea
as the so-called (nonequilibrium) spectral density function of the dot. 
Here, $G^{r/a}_d(\epsilon)$ are the retarded ($r$) and advanced ($a$) Green's functions. 
In general, for an interacting system, the spectral function depends on the electronic-bath parameters,
namely, the chemical potential $\mu_{\alpha}$ and the temperature $T_{\alpha}$, in a non-trivial way. 
The causality condition for the retarded and the advanced  Green's functions ensures the 
sum-rule for the spectral function, 
$\int _{-\infty}^{\infty} \frac{d\epsilon}{2\pi}  A(\epsilon; \{\mu_{\alpha},T_{\alpha}\})=1$.

A formal expression for $A(\epsilon; \{\mu_{\alpha},T_{\alpha}\})$ can be organized from the Dyson's equation 
\cite{bijay-wang-review, Haug-book, Rammer_review, Rammer_book} 
for the  interacting Green's function,
\be
G_d^{r/a}(\epsilon) = G_{0,d}^{r/a}(\epsilon) + G_{0,d}^{r/a}(\epsilon)\, \Sigma_n^{r/a}(\epsilon)\,G_d^{r/a}(\epsilon).
\label{dyson-eq}
\ee
Here, $G_{0,d}^{r/a}(\epsilon)$ is the Green's function for the dot, evaluated with the 
quadratic part of the Hamiltonian, $H_0$.
$\Sigma_n^{r/a}(\epsilon)$ is the nonlinear self-energy component, 
arising from many-body interactions on the dot.
Alternatively, we can write Eq.~(\ref{dyson-eq}) as 
\be
\Big[G_d^{r/a}(\epsilon)\Big]^{-1} = \Big[G_{0,d}^{r/a}(\epsilon)\Big]^{-1} - \Sigma_n^{r/a}(\epsilon).
\ee
This form allows us to write down the following expression for the spectral function,
\be
 A(\epsilon; \{\mu_{\alpha},T_{\alpha}\}) = 2 \, G_d^r(\epsilon) 
\left[ \Gamma(\epsilon) + \Gamma_n(\epsilon) /2 \right] \, G_d^a(\epsilon),
\label{eq:spectral}
\ee
with the many-body self energy
\bea
\Gamma_n(\epsilon) \equiv i \left[\Sigma_n^r(\epsilon) -\Sigma_n^a(\epsilon)\right]= 
i \left[\Sigma_n^>(\epsilon) -\Sigma_n^<(\epsilon)\right].
\eea
This function is responsible for broadening the spectral function of electrons---on top of the 
broadening from the metal-impurity hybridization.
Note that, in general, temperature and bias may affect all components  of the interacting Green's function.

Next, we write down an expression for the averaged electronic population on the dot. 
It can be related to the lesser component of the interacting Green's function, i.e.,
\bea
\langle n_d \rangle &\equiv& \langle d^{\dagger} d \rangle = -i \int_{-\infty}^{\infty} \frac{d\epsilon}{2 \pi} G_d^<(\epsilon) \nonumber \\
&=&-i \int_{-\infty}^{\infty}  \,\frac{d\epsilon}{2 \pi} \, G_d^r(\epsilon) \Sigma_{tot}^{<}(\epsilon) G_d^a(\epsilon).
\label{eq:navg}
\eea
The total self energy
$\Sigma_{tot}^{<}(\epsilon)= \Sigma_L^{<}(\epsilon)+ \Sigma_R^{<}(\epsilon)+\Sigma_n^{<}(\epsilon)$ 
includes contributions from the leads and from many-body interactions on the dot.
In the second line of the above equation we employ the Keldysh equation 
\cite{Kadanoff, Keldysh, book1,bijay-wang-review} for the lesser component $G_d^{<}(\epsilon)$.

Taking the metal-molecule hybridization $\Gamma$ to be a constant, independent of energy,
simple algebraic manipulations of Eqs.~(\ref{eq:Lan-cur}) and (\ref{eq:navg}) result in
the following relation between the charge current and the average dot occupation,
\begin{widetext}
\be
\langle I_L \rangle =
e \Gamma \Big[\langle n_d\rangle
-\int_{-\infty}^{\infty} \frac{d\epsilon}{2\pi } 
\big[ G_d^r(\epsilon) \, \Big(-i \Sigma_n^{<}(\epsilon) - \frac{f_L(\epsilon) + f_R(\epsilon)}{2} \Gamma_n(\epsilon) \Big)\, G_d^a(\epsilon) \big] 
- \int_{-\infty}^{\infty} \frac{d\epsilon}{2\pi }   A(\epsilon; \{\mu_{\alpha},T_{\alpha}\})  f_R(\epsilon) \Big].
\label{eq:central-eq}
\ee
\end{widetext}
This exact result holds only for symmetric junctions with constant hybridization $\Gamma$---yet 
irrespective of the nature and strength of many-body interactions on the dot. 

The discussion so far is rather standard  \cite{Haug-book},
yet it is illuminating to examine the different terms in Eq. (\ref{eq:central-eq}).
From this relation, it appears as if the departure of the
charge current from the average site population depends on two factors: (i) The deviation from the equilibrium 
fluctuation-dissipation theorem for the nonlinear self-energy component, 
as reflected by the second term in the R.H.S of Eq.~(\ref{eq:central-eq}). (ii) 
The integral over the spectral function of the interacting system, weighted by the right-lead distribution.
However, one can show that the first contribution vanishes identically 
under both equilibrium and nonequilibrium conditions, based on two fundamental principles.  

First, at equilibrium,  $\mu_L\!=\!\mu_R\!=\!\mu$ and $T_L\!=\!T_R\!=\!T$,  $f(\epsilon)=f_{\alpha}(\epsilon)$,
the detailed balance condition between the lesser and greater components of the Green's functions 
$G_d^{>}(\epsilon) = -e^{\beta(\epsilon-\mu)} \, G_d^{<}(\epsilon)$ ensures the fluctuation-dissipation relation for the nonlinear self-energy, 
\be
\Sigma_n^{<}(\epsilon) = -f(\epsilon) \left[\Sigma_n^r(\epsilon) - \Sigma_n^a(\epsilon)\right]
=i f(\epsilon) \Gamma_n(\epsilon).
\ee
As a result, the second term in the R.H.S of Eq.~(\ref{eq:central-eq}) vanishes, 
and we recover the standard expression for the average 
site population, in equilibrium, $\langle n_d \rangle^{eq} = \int \frac{d\epsilon}{2\pi } A(\epsilon, \{\mu, T\})f(\epsilon)$. 

What happens under general nonequilibrium conditions ($\mu_L \!\neq\! \mu_R$ and $T_L \neq T_R$)? 
In this case, current conservation (in steady state) imposes a constraint on the nonlinear self-energy, 
given as \cite{Haug-book}
\be
\!\!\int_{-\infty}^{\infty} \frac{d\epsilon}{2\pi} \, \Big[G_d^{>}(\epsilon) \Sigma_{n}^{<}(\epsilon) - G_d^{<}(\epsilon) \Sigma_{n}^{>}(\epsilon)\Big]=0.
\ee
This expression can be reorganized as 
\bea
\!\!\int_{-\infty}^{\infty} \!\!\frac{d\epsilon}{2\pi}\Big[\Big(G_d^{>}(\epsilon)\!&-&\!G_d^{<}(\epsilon)\Big) \Sigma_{n}^{<}(\epsilon) \nonumber \\
\!&-&\! G_d^{<}(\epsilon) \Big(\Sigma_{n}^{>}(\epsilon)\!-\!\Sigma_{n}^{<}(\epsilon)\Big)\Big]=0,
\nonumber\\
\eea
which implies that
\bea
\int_{-\infty}^{\infty} \frac{d\epsilon}{2\pi} \, 
\Big[A(\epsilon) \Sigma_{n}^{<}(\epsilon)-G_d^{<}(\epsilon) \Gamma_n(\epsilon)\Big]=0.
\eea
Plugging in the spectral function, Eq.~(\ref{eq:spectral}), we
receive the following condition,
\bea
&&\int_{-\infty}^{\infty} \frac{d\epsilon}{2\pi } \left[ G_d^r(\epsilon) \, \Big(-i \Sigma_n^{<}(\epsilon) - \frac{f_L(\epsilon) + f_R(\epsilon)}{2} \Gamma_n(\epsilon) \Big)\, G_d^a(\epsilon) \right]
\nonumber\\
&&=0.
\label{eq:cond}
\eea
This combination identically annihilates the second term in the R.H.S. of Eq.~(\ref{eq:central-eq}). 
Therefore, the relationship between the steady state charge current and the average electronic population, 
Eq.~(\ref{eq:central-eq}), in fact simplifies to 
\be
\langle I_L \rangle =e \Gamma \Big[\langle n_d\rangle-\int_{-\infty}^{\infty} \frac{d\epsilon}{2\pi }   A(\epsilon; \{\mu_{\alpha},T_{\alpha}\})  f_R(\epsilon) \Big].
\label{eq:cur-pop}
\ee
This expression can be also derived by imposing the current conservation condition at the starting point, in Eq.~(\ref{eq:MW}) \cite{Ness,NessC}.

Herewith, we will consider the bias dependence to enter only through the left-lead Fermi function. 
Few simple conclusions then immediately follow. 
First, if the spectral function is independent of bias 
(for example, if $H_n$ is ignored), we receive the following relation 
between the differential conductance $G(V)\equiv d\langle I_L \rangle/{dV}$ 
and the charge susceptibility $\chi_d(V) \equiv d\langle n_d \rangle/{dV}$, 
\be
G(V) = e \, \Gamma \,\chi_d(V).
\ee 
Moreover, in the limit $\mu_R \to -\infty$ the second term in the R.H.S of Eq.~(\ref{eq:cur-pop}) vanishes, and we arrive 
at the following equality, valid for an {\it arbitrarily  interacting} system, 
\be
\langle I_L \rangle =e \Gamma \langle n_d\rangle.
\label{eq:high-bias}
\ee
This seemingly trivial, steady state current-occupation
relation is the focus of our discussion throughout the rest of the paper.
Assumptions involved are that the junction is symmetric, $\Gamma(\epsilon)$ is constant,
and that the current is uni-directional, left to right. 
%
%
Why is Eq. (\ref{eq:high-bias}) interesting?
Typically, to examine convergence one {\it separately} studies the approach of the current and occupation
to fixed values. However, this relation offers an additional test for evaluating simulation tools.
In the next sections,
we test this equality on the Anderson-Holstein model and the electron-interacting Anderson impurity model.
We demonstrate below that by considering Eq. (\ref{eq:high-bias}), we identify inconsistencies in simulation techniques,
resulting in e.g. spurious transport phenomena.
Another reason for simplifying Eq. (\ref{eq:cur-pop}), and using the specific limit (\ref{eq:high-bias}) is that in many techniques, as used below, 
one does not have a direct access to the impurity spectral function  $A(\epsilon; \{\mu_{\alpha},T_{\alpha}\})$, rather, only the current and the dot occupation 
can be computed.

Before turning to particular models, we recall that phase-loss and inelastic effects can be introduced into a
noninteracting transport behavior 
(under the Hamiltonian $H_0$) by attaching the dot to  ``Buttiker probes"  \cite{Buttiker}.  For recent applications
in the context of quantum dot and molecular transport junctions see  \cite{Salil1,Kilgour1,Kilgour2}.
Working with e.g. the dephasing probe technique, one can readily prove that in the case of a single impurity,
the method exactly satisfies Eq. (\ref{eq:cur-pop}). 
Similarly, a cheap approach for introducing inelastic effects,
by adding a broadening into the noninteracting transmission function,
satisfies Eq. (\ref{eq:high-bias}). For a recent theory-experiment study of 
coherent and incoherent effects in single-dot junctions
see \cite{Nijhuis}. 

\begin{center}
\begin{figure*}[t]
\includegraphics[width=20cm]{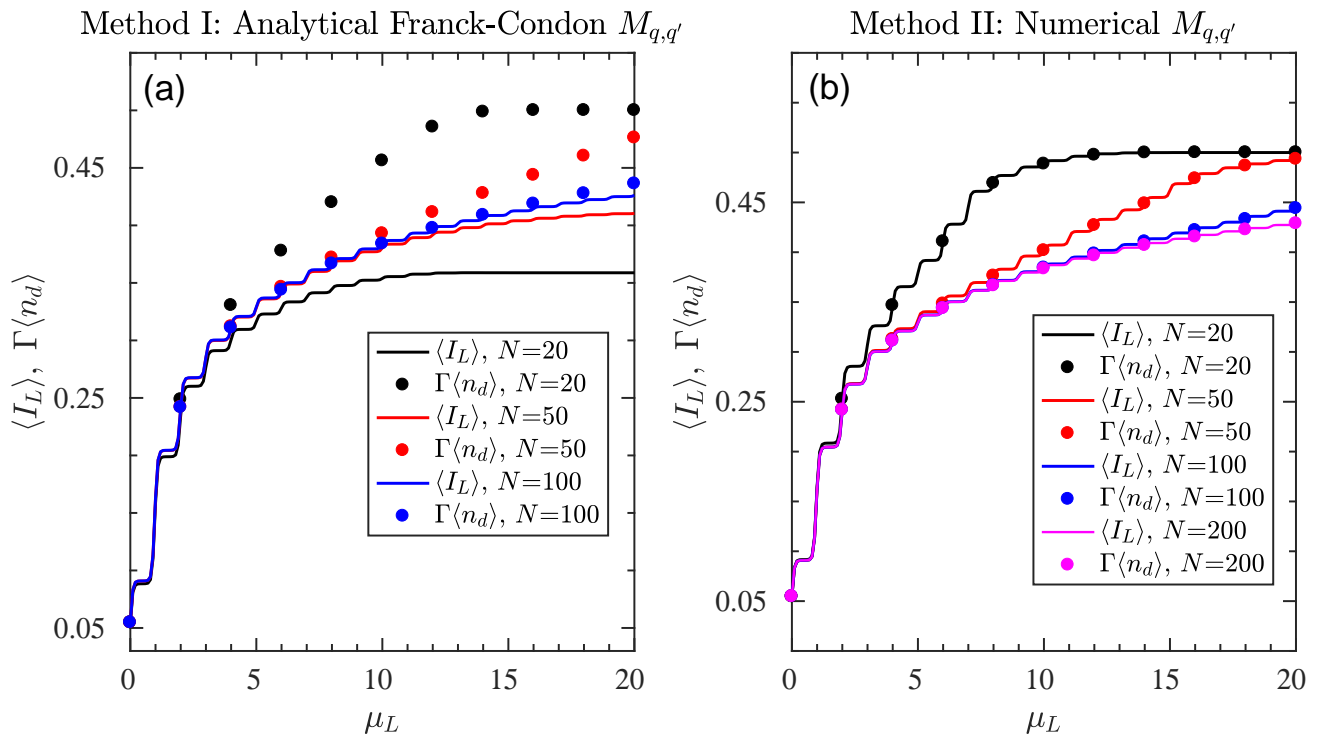}
\caption{
Steady state charge current $\langle I_L\rangle$ (solid) 
and average population $\Gamma \langle n_d \rangle$ ($\circ$) as a function of 
left-lead chemical potential $\mu_L$ for strong electron-phonon coupling strength $\gamma_0/\omega_0=2$. 
Matrix elements $M_{q,q'}$ are calculated (a) using the Franck Condon formula, Eq. (\ref{eq:FC}),
(b) using  numerical diagonalization at finite $N$.
The number of levels included for the truncated harmonic oscillator are $N=20$ (black), $N=50$ (red), $N=100$ (blue),
and $N=200$ (magenta). The last case is demonstrated only with Method II, as we could not evaluate
high order FC factors with the analytical formula (\ref{eq:FC}).
Parameters are $\epsilon_d-\alpha_0^2\omega_0$=0, $\omega_0$ =1, $\gamma_0$=2,  
$\mu_R=-100$,  $T_L=T_R=0.05$.
}
\label{Fig1}
\end{figure*}
\end{center}

\begin{center}
\begin{figure*}
\includegraphics[width=19.5cm]{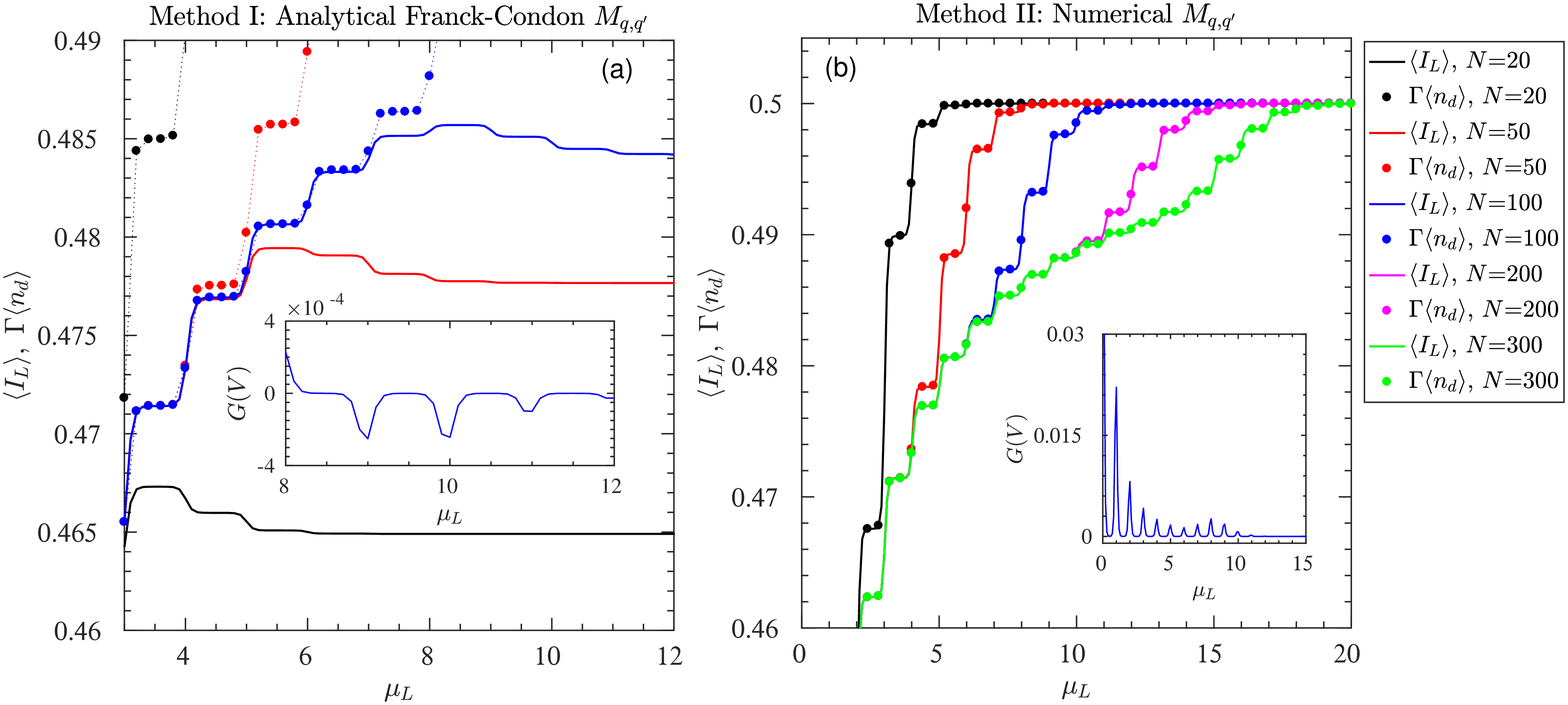} 
\caption{
Steady state charge current and average population as a function of left-lead chemical potential $\mu_L$
for intermediate electron-phonon coupling, $\gamma_0/\omega_0=0.5$. 
Matrix elements $M_{q,q'}$ are calculated (a) using the Franck Condon formula, Eq. (\ref{eq:FC}),
(b) using numerical diagonalization at finite $N$.
$N=20$ (black), $N=50$ (red), $N=100$ (blue), $N=200$ (magenta), $N=300$ (green).
The last two values are only demonstrated using Method II.
The insets display the differential conductance $G(V)$ for $N=100$. 
Other parameters are the same as in Fig.~(\ref{Fig1}).}
\label{Fig2}\end{figure*}
\end{center}

\section{Anderson-Holstein Model: polaronic QME}
\label{SAH}



As our first example, we consider the spinless Anderson-Holstein model \cite{AHcomment}
in which electrons on  the dot are coupled 
to a single bosonic (photonic or phononic) mode, creation and annihilation operators $b^{\dagger}$, $b$.
For simplicity, in this example we do not include e-e repulsion, but it is trivial to consider that. 
The Hamiltonian of the dot, along with the bosonic component is
\be
H_{d0}+H_n= \epsilon_d \, n_d + \omega_0 \, b^{\dagger} b+ \gamma_0\, n_d\, (b+b^{\dagger}).
\label{eq:AH}
\ee
Recall that $\epsilon_d$ is the electronic site energy, $n_d =d^{\dagger} d$ is the number operator for the dot electrons.
Other parameters are $\omega_0$ as the frequency of the bosonic mode and $\gamma_0$ as the electron-boson coupling energy. 
This model is often examined in the context of phononic-vibrational effects in molecular electronic
junctions \cite{Galperin-review}. It can be realized as well in circuit-QED experiments \cite{Kontos-cQED, LeHur-cQED}.
In what follows, we refer to the boson mode as a ``phonon".

In the limit of weak electron-phonon coupling (single boson excitation/de-excitation), 
we only include the lowest non-zero order (the second order) in the nonlinear electronic self-energy 
 $\Sigma_{n}^{<,>,r,a}(\omega)$.
It can be obtained by following the standard Hartree and Fock-like diagrams \cite{Galperin-review}. 
The Hartree term trivially satisfies Eq.~(\ref{eq:cond}): Since it is local in contour time, 
$\Sigma_{H}^{</>}(\omega)=0$. 
The Fock diagram, in contrast, has non-zero lesser and greater components,  but it satisfies 
the condition (\ref{eq:cond}). Therefore, we conclude that Eq.~(\ref{eq:cur-pop}) [and obviously Eq.~(\ref{eq:high-bias})]
hold under a perturbative treatment at the Hartree-Fock level. 
 
Next, we consider the strong electron-phonon coupling limit, and test Eq. (\ref{eq:high-bias}). We use 
the polaronic quantum master equation (QME) approach, as employed e.g. in
Refs.~\cite{Mitra,KochSela,Koch1,Koch2,Wege,Thoss11a, Thoss11b,Simine,Gernot1,Gernot2}. 
We briefly review the principles of this approach.
To examine the strong coupling limit, one first performs the small-polaron transformation so as to eliminate
the electron-phonon interaction term from within $H_n$. As a result of the transformation,
the dot energy is renormalized, $\epsilon_d \to \epsilon_d - \gamma_0^2/\omega_0$, 
while the dot-metal tunneling term is dressed by the translational operator 
$D(\alpha_0)= e^{-\alpha_0 (b^{\dagger}-b)}$, where $\alpha_0=\frac{\gamma_0}{\omega_0}$ is a dimensionless
electron-phonon interaction parameter, $c^{\dagger}_{k\alpha}d \rightarrow D(\alpha_0)c^{\dagger}_{k\alpha}d $

A kinetic quantum master equation can be derived by following standard approximations,  namely, 
the secular Born-Markov approximation, assuming (i) weak metal-dot coupling $\Gamma<T,\Delta \mu$,
(ii) fast decay of the baths' (metals) correlation functions vs.
slow tunneling dynamics, (iii) fast decay of vibrational coherences.
While this equation captures only lowest-order processes in the tunnelling energy,
it treats the electron-phonon interaction exactly---in that order of the tunnelling coupling.

We define a reduced density matrix (RDM) $\rho_s$ for the interacting impurity, which includes the dot and the local phonon.
The diagonal elements of the RDM,
$p^{n}_{q}(t) \equiv \langle n,q| \rho_s(t)|n, q\rangle$, satisfy kinetic-like equations of motion
%
\be
\dot{p}^{n}_{q} =  \sum_{n', q'} \Big( p_{q'}^{n'} k_{q' \to q}^{n' \to n} - p_{q}^{n} \, k_{q \to q'}^{n \to n'}\Big).
\label{eq:pop}
\ee
Here, $n=0,1$ represents an empty or occupied electronic site and $q$ identifies the 
state of the phonon mode $(q = 0, 1, 2, \cdots)$.  
Once steady state is achieved, $\dot{p}_q^{n,ss}=0$, 
the averaged charge current and population
can be computed from the following expressions ($\alpha=L,R$),
\bea
\langle I_{\alpha} \rangle &=& \sum_{q, q'}  \Big( k^{0 \to 1}_{q \to q', \alpha} \, p_q^{0,ss} - k^{1 \to 0}_{q \to q', \alpha} \, p_q^{1,ss} \Big), 
\label{eq:AHcur}
\\
\langle n_d \rangle &=& \sum_{q} p_{q}^{1,ss}.
\label{eq:AHn}
\eea
For convenience, we set $e=1$.
In Eq.~(\ref{eq:pop}), $k_{q \to q'}^{n \to n'}=\sum_{\alpha} k_{q \to q',\alpha }^{n \to n'}$ 
is the total rate constant for the transition $|n, q\rangle \to |n', q'\rangle$. 
It is easy to verify that this scheme conserves current, $\langle I_L\rangle + \langle I_R\rangle =0$.
Note that according to our sign convention, the current is positive when flowing towards the dot.
The rate constants satisfy
\bea
k_{q \to q', \alpha }^{ 0 \to 1} &=& \Gamma f_{\alpha} \big(\epsilon_d - \frac{\gamma_0^2}{\omega_0} + \omega_0 (q'-q)\big) \big|M_{q,q'}\big|^2 \nonumber \\ 
k_{q \to q', \alpha }^{1 \to 0} &=& \Gamma \Big[1-f_{\alpha} \big(\epsilon_d - \frac{\gamma_0^2}{\omega_0} + \omega_0 (q'-q)\big)\big] \big|M_{q,q'}\big|^2,
\nonumber\\
\label{eq:rates}
\eea
%
where in these expressions, the Fermi function is calculated at the renormalized energy, with a certain number of quanta 
$\omega_0$, and
%
\bea
M_{q,q'} = \langle q| e^{-\frac{\gamma_0}{\omega_0} (b^{\dagger}-b)} | q'\rangle,
\eea
are the matrix elements of the shift operator. For a harmonic (boson) mode, 
these are the familiar Franck-Condon (FC) factors.
In terms of the dimensionless parameter $\alpha_0$, the FC factors are given by,
\bea
M_{q,q'} &\equiv&\langle q |  e^{-\alpha_0(b^{\dagger}-b)}|q'\rangle, \,\,\,\,\, q,q'=0,1,2... 
\nonumber\\
&=& sign(q'-q)^{q-q'}\alpha_0^{q_M-q_m}e^{-\alpha_0^2/2}\sqrt{\frac{q_m!}{q_M!}} L_{q_m}^{q_M-q_m}(\alpha_0^2),
\nonumber\\
\label{eq:FC}
\eea
with $q_m=\min\{q,q'\}$,  $q_M=\max\{q,q'\}$, and $L_{a}^{b}(x)$ as the generalized Laguerre polynomials.

It is straightforward to prove that the polaronic quantum master equation as detailed here fulfills 
the exact relation (\ref{eq:high-bias}). 
This fact is not obvious, given that several approximations are involved in the derivation, potentially 
defecting exact relations.
We begin with Eq.~(\ref{eq:AHcur}). In the limit
$\mu_R \to -\infty$, right-lead induced rates simplify as follows: $k_{q \to q', R }^{ 0 \to 1}=0$,
 and $k_{q \to q', R }^{1 \to 0} \approx \Gamma\, |M_{q,q'}|^2$.
The current at the right contact becomes
\bea
\langle I_R \rangle = -\Gamma\, \sum_{q} p_{q}^{1,ss}\, \sum_{q'} |M_{q,q'}|^2.
\label{eq:proof}
\eea
Next, we use the completeness relation for the bosonic manifold, $\sum_{q=0}^{\infty} |q \rangle \langle q| = I$,
with $I$ as the identity operator, the unitarity of the polaron transformation, $MM^{\dagger}=M^{\dagger}M=I$, 
and current conservation,
and get $\langle I_L \rangle = \Gamma\ \sum_q p_{q}^{1,ss} $, proving Eq. (\ref{eq:high-bias}).

In Fig. \ref{Fig1}, we present numerical simulations of the average current and the dot's population, 
based on Eqs. (\ref{eq:AHcur})-(\ref{eq:AHn}).
Obviously, in practice we truncate the harmonic spectrum, to include
a finite number $N$ of basis states for the mode. The particular choice of $N$
depends on model parameters: the applied
bias voltage, temperature, mode frequency, and the many-body interaction strength. 
In panel (a), the matrix elements in Eq. (\ref{eq:rates}) are evaluated 
with the analytical form for the FC factors, Eq. (\ref{eq:FC}). We refer to this approach as ``Method I".
Surprisingly, we observe a clear violation of the {\it exact} relation (\ref{eq:high-bias}).
This is unsettling since we had just {\it proved}
that formally, the polaronic QME technique should satisfy this relation. What is wrong then?
The answer is subtle---yet with significant ramifications. 
It is inconsistent to calculate the matrix elements $M_{q,q'}$ with the exact-analytical FC formula---yet work with 
a finite (even if large) basis for the mode $|q\rangle$. 

How can we rectify this, to satisfy Eq. (\ref{eq:high-bias}) in numerical simulations? 
Since we perform a summation over a finite basis in Eq. (\ref{eq:AHcur}),
we must calculate $M_{q,q'}$ in that finite basis,
without relying on the analytical (infinite-$N$) expression for the FC factors. This procedure,  ``Method II",
is explained in the Appendix, see also Ref. \cite{Simine}.
In fact, Method II, which is fully numerical, is very efficient:
While in Method I 
large factorials and high-order Laguerre polynomials should be calculated for high quantum states.
the diagonalization scheme (Method II) described in the Appendix provides all matrix elements $M_{q,q'}$
in a single operation. 
Using this method, we demonstrate in Fig. \ref{Fig1}(b) that the  current-occupation relation is exactly satisfied,
even before convergence to the exact ($N\rightarrow \infty$) limit is reached.

In Fig. \ref{Fig2} we illustrate another critical deficiency of Method I,
using an intermediate electron-phonon coupling strength. 
In panel (a) we observe that Method I not only violates  the exact 
relation (\ref{eq:high-bias}), it further leads to a fundamentally incorrect function, namely, 
negative differential conductance (NDC). This spurious effect, highlighted in the inset of panel (a),
survives even when the basis set is large, $N=100$. 
Beyond that, we are not able to compute higher order FC factors based on Eq. (\ref{eq:FC}).
In contrast, in panel (b) we show that Method II 
does not manifest the NDC phenomenon for any choice of $N$, even when few levels are employed.
Furthermore, Method II can be used to calculate the behavior of the junction
under very high voltages, when many vibrational states should be taken into account.


To conclude this discussion: One could argue that for large enough $N$, Eq. (\ref{eq:high-bias}) 
should be satisfied numerically---even when the FC factors
are evaluated analytically based on the harmonic spectrum.
However, our simulations demonstrate that Method I is prone to 
fundamental errors, and the fully-numerical Method II is superior is several ways:
(i) It identically satisfies the exact relation (\ref{eq:high-bias}).
(ii) It is highly efficient as one obtains all matrix elements $M_{q,q'}$ in a single operation.
(iii)  Method I may yield  incorrect features, even when many (seemingly sufficient) states are included,
as we had demonstrated in Fig. \ref{Fig2}.


\begin{figure*}
\includegraphics[width=17cm]{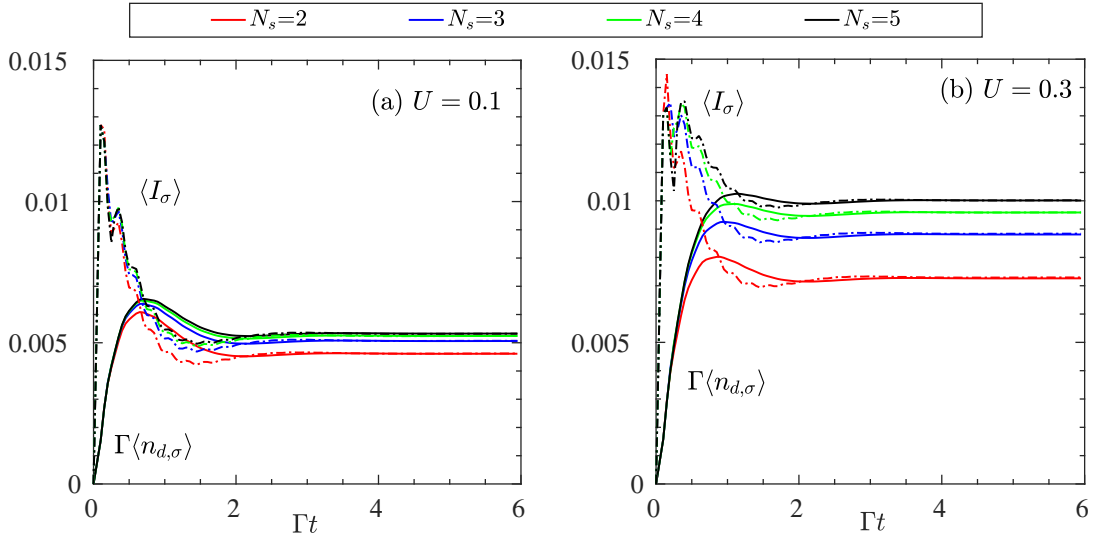}         
\caption{
Anderson model with an onsite electron-electron interaction.
Transient dynamics for the charge current (dashed) and dot population (full) for
(a) $U$=0.1 and (b) $U$=0.3.
Other parameters are $\delta t=1$, $D=\pm 1$, $\mu_L=0.2$, $\mu_R=-1$, $1/T_{L,R}=200$, $\epsilon_d+U/2=0.3$,
$\Gamma=0.05$, $L=200$ states per spin per bath.
Different lines correspond to different memory size, $N_s=2-5$, see the legend.
}
\label{Fig3}
\end{figure*}

\section{Interacting impurity model: Path integral simulations}
\label{SINFPI}

In this section, we test the  exact relation (\ref{eq:high-bias}) for 
the Anderson impurity model, including the interaction of spin-up and spin-down electrons on the dot.
We simulate the dynamics from a certain initial condition by employing a time-evolution scheme, 
an iterative, numerically-exact influence functional path integral approach.
The principles of this method, and its application to the interacting Anderson dot model, are detailed in Refs. \cite{IF1,IF2}.
Why is it nontrivial to verify Eq. (\ref{eq:high-bias}) with this tool?
In this numerically exact method (as well we in other numerical techniques), 
different observables are independently calculated directly from their definitions.
This should be contrasted e.g. to Green's function-based perturbative methods, where different transport observables are expressed
in terms of the spectral function, see Sec. \ref{SModel}. 
Given the centrality of the interacting Anderson impurity model 
for solving extended models through the dynamical mean field theory,
assessing accuracy of simulation tools is crucial.

We introduce the model. The dot Hamiltonian and the onsite interaction are 
\bea
H_{d0}+H_n= \sum_{\sigma} \epsilon_d n_{d,\sigma} + Un_{d,\uparrow}n_{d, \downarrow},
\eea
with $\sigma=\uparrow,\downarrow$, $n_{d,\sigma}=d^{\dagger}_{\sigma}d_{\sigma}$. $U$ denotes the e-e interaction strength.
The total Hamiltonian reads
\bea
H &=&  
\sum_{\sigma} \epsilon_d n_{d,\sigma} + Un_{d,\uparrow}n_{d, \downarrow}
\nonumber\\
&+& \sum_{\alpha,k,\sigma}\epsilon_{k\alpha\sigma} c_{k\alpha\sigma}^{\dagger} c_{k\alpha\sigma}
+\sum_{\alpha,k,\sigma} v_{k\alpha\sigma} c_{k\alpha\sigma}^{\dagger}d_{\sigma} + h.c. 
\nonumber\\
\eea
We assume spin degeneracy, and set a symmetric junction 
with the metal-dot hybridization energy
$\Gamma= 2\pi \sum_{k}|v_{k\alpha\sigma}|^2\delta(\epsilon-\epsilon_{k\alpha\sigma})$.
The two metals are described by flat bands with sharp cutoffs at $D=\pm 1$.
At the initial time, the metals are prepared in a nonequilibrium condition, with $\mu_L=0.2$ and $\mu_R=-1$.
The dot is empty at the beginning of our simulation.

We follow, separately, the real-time dynamics of
the dot occupation and the current with
a deterministic numerically exact approach, 
an influence functional path integral (INFPI) technique, described in details in Refs.  \cite{IF1,IF2}.
This method relies on the observation that in out-of-equilibrium (and/or finite
temperature) situations, bath correlations have a finite range,
allowing for their truncation beyond a memory time dictated
by the voltage bias and the temperature \cite{Makri1}. 
As convergence is facilitated at large bias, the method is perfectly suited for testing Eq. (\ref{eq:high-bias}).


As elaborated in Refs.  \cite{IF1,IF2}, INFPI suffers from several sources of numerical errors. 
First, the metals are discretized to 
include finite number of electronic states. 
We find here that fora band extending $D=\pm 1$, $L=200$ states
are sufficient to describe a seemingly irreversible dynamics up to $\Gamma t\sim 5$.
Other sources of error are the Trotter error,
given the approximate factorization of the 
short time evolution operator into quadratic and
many-body terms, and the memory error, resulting from the
truncation of the influence functional. Convergence is verified by demonstrating
that results are insensitive to the time step $\delta t$ and the memory
size $\tau_c=N_s\delta t$, with $N_s$ an integer.


In Fig.~(\ref{Fig3}) we display the transient dynamics for a certain spin species, $\langle I_{\sigma}\rangle$  and 
$\Gamma \langle n_{d,\sigma} \rangle$, for two different values of the interaction strength $U$. 
For simplicity, we set $e=1$.
The current displayed is the average at the two contacts, $\langle I_{\sigma}\rangle = \langle I_{L\sigma} - I_{R\sigma}\rangle/2 $.
It is remarkable to observe that an excellent agreement between these expectation values 
is obtained as we approach the long time limit, 
even before convergence to the exact limit ($N_s\rightarrow \infty$, $\delta t\rightarrow 0$) is achieved, see panel (b).
We can rationalize this agreement as follows. 
The INFPI method is deterministic, and it conserves the trace of the density matrix. 
For a given choice $N_s$, there are $2^{2N_s}$ paths that should be summed over, covering all the configurations
of the fictitious spin (received from the Hubbard-Stratonovich transformation). 
By simulating the population and current while
keeping all paths for a given $N_s$ and $\delta t$, we essentially maintain the dimension of 
the ``Hilbert space" of the problem. As a result, we
satisfy the exact relation (\ref{eq:high-bias})---even when simulation results are away from the exact answer.
Small deviations observed in Fig. \ref{Fig3} arise from (i) not arriving yet at the (quasi) steady state limit,  (ii)
representing the leads with finite number of states, (iii) using parameters deviating from the
assumptions behind Eq. (\ref{eq:high-bias}): $D=\pm1$ and $\mu_R=-1$.

We argue that it is important to validate numerically exact methods
by verifying (\ref{eq:high-bias}) in steady state. The fact that INFPI preserves this relation even before convergence 
allows us to regard simulation results for $\langle n_d \rangle$ and $\langle I\rangle$ 
as physically-meaningful values for the true population and current.
Techniques based on sampling and filtering paths are promising avenues for accelerating path integral simulations 
\cite{Makri2,Makri3}.
Nevertheless, these methods break the exact current-population relation, and one would need
to affirm convergence by verifying Eq. \ref{eq:high-bias}.


\section{Summary}
\label{Ssum}

By employing the nonequilibrium Green's function approach, we 
had established an exact relationship between the steady state electronic current 
and the average electronic occupation for a symmetric, interacting single impurity junction. 
We had tested this relation in two cases, for the Anderson Holstein model, and for the e-e interacting Anderson dot model,
using perturbative and numerically exact approaches, respectively. 
 
For the Anderson Holstein model, testing the relation had assisted us in uncovering a subtle inconsistency in the
popular polaron QME numerical implementation---with important ramifications for clearing out spurious transport effects.
In the case of the electron-interacting Anderson dot model we found that a deterministic (trace conserving)
influence functional path integral approach obeys the relation
even before convergence to the exact-asymptotic limit is reached. 

Simulating the Anderson dot model is a nontrivial task. 
Testing Eq. (\ref{eq:high-bias}) is suggested here as a viable mean for carefully scrutinizing
the consistency and accuracy of numerical schemes. 
Other consistency checks involve the study of transport symmetries, e.g. the  Onsager theorem under
perturbative methods \cite{Wacker}.
Finally, if full-counting statistics information is available,
one could probe the validity of approximate tools 
by comparing high order cumulants \cite{bijay-recon}, 
or more fundamentally, by confirming that the nonequilibrium fluctuation symmetry is satisfied \cite{bijay-vibration}.


We acknowledge Herve Ness for helpful discussions.
The work of DS and BKA was supported by the Natural Sciences and Engineering Research Council of Canada, 
the Canada Research Chair Program, 
and the Centre for Quantum Information and Quantum Control (CQIQC) at the University of Toronto.


\renewcommand{\theequation}{A\arabic{equation}}
\setcounter{equation}{0}  

\section*{Appendix: Anderson dot model coupled to an $N$-level system}

We describe here ``Method II", a fully numerical scheme for calculating current and population in the 
Anderson-Holstein model at the level of the polaronic rate equation.
In this model, electrons on the dot interact with a local bosonic mode, (\ref{eq:AH}).
Here we consider a more general construction, with the dot coupled to an $N$-level system. The interacting (mode 
displacement) operator is similarly represented by a finite matrix.
For completeness, we recount the different terms in the total Hamiltonian
\bea
H= H_{d0}+ H_L+H_R + H_{dL}+H_{dR}+ H_n.
\eea
The dot interacts with two metals,
$H_{\alpha} =\sum_{k}\epsilon_{k\alpha} c_{k\alpha}^{\dagger}c_{k\alpha}$
with the tunneling Hamiltonian
\bea
H_{dL}+H_{dR}=\sum_{\alpha,k}\left(v_{k\alpha}c_{k\alpha}^{\dagger}d 
+v_{k\alpha}^*d^{\dagger}c_{k\alpha}\right).
\label{eq:Apphyb}
\eea
The Hamiltonian of the dot and the $N$-level entity is
\bea
H_{d0}+H_n&=& \epsilon_d n_d 
\nonumber\\
&+&
\sum_{q=0}^{N-1}\epsilon_q |q\rangle \langle q| + \alpha_0 n_d \sum_{q,q'}F_{q,q'}|q\rangle \langle q'|.
\eea
As before, $n_d=d^{\dagger}d$ denotes the electron occupation number operator for the dot.
The Hamiltonian of the $N$-level system is written in its energy representation
with states $|q\rangle$, $q,q'=0,1,...,N-1$.
This finite system is coupled to the electron number operator via the operator $F$,
$\alpha_0$ is a dimensionless parameter.
It is useful to define the $N$-level Hamiltonian separately, 
\bea
H_{N}=\sum_q\epsilon_q|q\rangle \langle q|
+\alpha_0\sum_{q,q'}F_{q,q'}|q\rangle\langle q'|.
\label{eq:Himp}
\eea
This operator is hermitian, and it can be diagonalized with a unitary transformation
\bea
\bar H_{N}= e^AH_{N} e^{-A},
\label{eq:U}
\eea
where $A^{\dagger}=-A$ is an anti-hermitian operator.
We now introduce a related unitary operator, $e^{A n_d}$. Note that
$e^{A n_d}de^{-A n_d}=de^{-A}$ and that $e^{A n_d}d^{\dagger}e^{-A n_d}=d^{\dagger}e^{A}$.
Operating on the original, total Hamiltonian, $\bar H=e^{A n_d}He^{-A n_d}$, we get
\bea
\bar H&=&\sum_{\alpha=L,R,k}\epsilon_{k\alpha}c_{k\alpha}^{\dagger}c_{k\alpha} + \epsilon_d  n_d
\nonumber\\
&+&
\sum_{\alpha,k}\left(v_{k\alpha}c_{k\alpha}^{\dagger}de^{-A} +v_{k\alpha}^*d^{\dagger}c_{k\alpha}e^{A}\right)
\nonumber\\
&+& (1- n_d)\sum_q\epsilon_{q}|q\rangle\langle q| + n_d \bar H_{N}.
\label{eq:barH}
\eea
%

To simulate the Anderson-Holstein model with a local harmonic  mode linearly coupled to the dot
we assume a certain-large $N$ (such that $N\omega_0 \gg  \Delta\mu$, $T$). We
set $\epsilon_q=q\omega_0$ and
discretize the bosonic displacement operator $b^{\dagger}+b$,
\bea
F_{q,q'}=\omega_0\sum_{q,q'}\sqrt q|q\rangle \langle q'|\delta_{q'=q-1} +h.c.
\label{eq:Fppp}
\eea
Here $\delta_{q,q'}$ is the Kronecker delta. 
In the limit $N\rightarrow \infty$, 
$A=\alpha_0(b^{\dagger}-b)$ and $\bar H_{N}= \omega_0b^{\dagger}b-\alpha_0^2\omega_0$. 
Therefore, in this limit we can write down the polaronic Hamiltonian as 
\bea
\bar{H} 
&=&
\sum_{\alpha,k}\epsilon_{k\alpha} c_{k\alpha}^{\dagger}c_{k\alpha} +\epsilon_d  n_d
\nonumber\\
&+& \sum_{\alpha,k}\left[v_{k\alpha}c^{\dagger}_{k\alpha}de^{-A} +v_{k\alpha}^*d^{\dagger}c_{k\alpha}e^{A} \right]
\nonumber\\
&+&  \sum_{q}q\omega_0|q\rangle \langle q| - \alpha_0^2 \omega_0 n_d.
\label{eq:barHAH}
\eea
This Hamiltonian is the starting point for our ``Method II" QME calculations. 
We use {\it finite} number of levels $N$,
with $e^A$ obtained from the diagonalization of $H_N$ in Eq. (\ref{eq:Himp}).
Isn't there an inconsistency in this approach? We calculate $e^A$ based on a finite representation for the local mode, yet
 still assume that its spectrum after diagonalization follows the harmonic-like behavior.
%
Indeed, if we back-transform,
$e^{-A} \left[\sum_q \omega_0q |q\rangle \langle q| -\alpha_0^2\omega_0\right]e^{A}$, 
for a finite $N$, we will not receive Eq. (\ref{eq:Himp}) identically with (\ref{eq:Fppp}), rather, we will get
a mode slightly deviating from the harmonic spectrum.
What is crucial though for satisfying Eq. (\ref{eq:high-bias}) is that
matrix elements $M_{q,q'}$ are calculated with a complete-finite basis.
%

Summing up, in method II we put together an $N\times N$-sized matrix $H_N$ (\ref{eq:Himp}) with the discretized 
displacement operator (\ref{eq:Fppp}), and diagonalize it.
The elements of the transformation matrix $M_{q,q'}=\langle q|e^A|q'\rangle$
correspond to the Franck-Condon factors of Eq. (\ref{eq:FC}). 
They are used to calculate the rate constants (\ref{eq:rates}), then current and population
following Eq. (\ref{eq:AHcur})-(\ref{eq:AHn}).

It is important to note that both Methods, I and II, are defective:
In Method I, we use a truncated harmonic spectrum, but with FC factors evaluated based on an infinite harmonic manifold.
In Method II,  the matrix elements corresponding to the FC factors
are received from the numerical diagonalization of a finite matrix, 
but the energies of the $N$-level mode are taken equi-distant from a perfect harmonic spectrum.
In both techniques, the asymptotic limit $N\rightarrow \infty$ provides the exact result. 
While both approaches are inexact, 
Method II is superior since it preserves the charge-occupation symmetry.



\begin{thebibliography}{0}
\bibitem{NessC}
It is interesting to note that based on Eq. (\ref{eq:Lan-cur}), one can immediately organize Eq. (\ref{eq:cur-pop}) in a similar form,
%
\be
\langle I_L \rangle =-e \Gamma \Big[\langle n_d\rangle - \int_{-\infty}^{\infty} \frac{d\epsilon}{2\pi }   A(\epsilon; \{\mu_{\alpha},T_{\alpha}\})  f_L(\epsilon) \Big].
\label{eq:cur-pop2}
\ee
This expression was the starting point in previous studies \cite{Ness} over the relation between the charge current and the dot occupation.
\bibitem{AHcomment}
One can account for the spin degree of freedom
by revising the rate constants (\ref{eq:rates}) to include prefactors, 
$s(n=0\to n=1)=2$ and $s(n=1\to n=0)=1$.
This modification does not affect any of our observations.
\end{thebibliography}
\end{document}